\documentclass[journal]{IEEEtran}
\usepackage[utf8]{inputenc}
\usepackage[noadjust]{cite}

\usepackage{pgf, pgfplots, tikz}
\pgfplotsset{compat=newest}
\usetikzlibrary{shapes.geometric, arrows}
\usetikzlibrary{positioning}
\usetikzlibrary{plotmarks}
\usetikzlibrary{arrows.meta}
\usepgfplotslibrary{patchplots}
\usepackage{grffile}
\newlength\fwidth
\newlength\fheight

\usepackage{subfig}

\usepackage{graphicx}
\usepackage{amsmath}
\usepackage{amssymb}
\usepackage{booktabs}
\usepackage{mathtools, nccmath}
\usepackage{algorithm,algorithmic}

\usepackage{diagbox}

\usepackage{amsmath}
\DeclareMathOperator*{\argmax}{arg\,max}

\DeclareMathOperator*{\angleF}{angle}
\DeclareMathOperator*{\real}{Re}

\begin{document}
\bstctlcite{IEEEexample:BSTcontrol}

\title{A Low-Complexity LoRa Synchronization Algorithm Robust to Sampling Time Offsets}

\author{Mathieu~Xhonneux,~\IEEEmembership{Graduate~Student~Member,~IEEE,}
        Orion~Afisiadis,~\IEEEmembership{Member,~IEEE,}
        David~Bol,~\IEEEmembership{Senior Member,~IEEE,}
        and~Jérôme~Louveaux,~\IEEEmembership{Member,~IEEE}
\thanks{Mathieu Xhonneux, David Bol and Jérôme Louveaux are with ICTEAM, UCLouvain, 1348 Louvain-la-Neuve, Belgium (e-mail: mathieu.xhonneux@uclouvain.be).}
\thanks{Orion Afisiadis is with the Telecommunication Circuits Laboratory, EPFL, 1015 Lausanne, Switzerland.}
\thanks{Copyright (c) 20xx IEEE. Personal use of this material is permitted. However, permission to use this material for any other purposes must be obtained from the IEEE by sending a request to pubs-permissions@ieee.org.
This document is the paper as accepted for publication in IEEE IoT Journal, the fully-edited paper is available at https://ieeexplore.ieee.org/document/9501038.
DOI: 10.1109/JIOT.2021.3101002}}

\maketitle

\medskip

\begin{abstract}
    LoRaWAN is nowadays one of the most popular protocols for low-power Internet-of-Things communications.
    Although its physical layer, namely LoRa, has been thoroughly studied in the literature,
    aspects related to the synchronization of LoRa receivers have received little attention so far.
    The estimation and correction of carrier frequency and sampling time offsets is however crucial to
    attain the low sensitivity levels offered by the LoRa spread-spectrum modulation.
    The goal of this paper is to build a low-complexity, yet efficient synchronization algorithm
    capable of correcting both offsets.
    To this end, a complete analytical model of a LoRa signal corrupted by these offsets is first derived.
    Using this model, we propose a new estimator for the sampling time offset.
    We also show that the estimations of the carrier frequency and the sampling time offsets cannot be performed independently.
    Therefore, to avoid a complex joint estimation of both offsets, an iterative low-complexity synchronization
    algorithm is proposed.
    To reach a packet error rate of $10^{-3}$, performance evaluations show that the proposed receiver requires only $1$ or $2$~dB higher SNR than
    a theoretical perfectly synchronized receiver, while incurring a very low computational overhead.
\end{abstract}


\section{Introduction}

In the last years, the rise of the Internet of Things (IoT) led to the emergence of new low-power wide-area network technologies.
The design of these new protocols was motivated by the proportionally high energy cost of wireless communications on low-power end nodes~\cite{bol-esscirc2018,bol-isscc2018,bol-springer2020},
and the high complexity of the 3GPP cellular standards.
Among the new IoT low-power standards, LoRaWAN has become one of the most popular and widely deployed solutions~\cite{mekki2019comparative}.
LoRaWAN consists of a protocol stack including a physical layer (PHY), usually called LoRa, and a MAC layer. 
The LoRa PHY layer is a proprietary standard patented by Semtech~\cite{seller2016patent,seller2018patent}, whereas the MAC layer is 
an open standard defined by the LoRa Alliance~\cite{sornin2015lorawan}.
Due to its proprietary nature, the only transceivers compatible with LoRa are commercialized in agreement with Semtech, and no precise specifications
of the PHY layer are publicly available.

Although the literature on LoRa has grown rapidly over the years~\cite{haxhibeqiri2018survey},
major aspects of its physical layer have not yet been studied. The principles of the modulation and demodulation stages are well-known
\cite{vangelista2017frequency,chiani2019lora}, but little research has been conducted on the synchronization stage.

The principal offsets to mitigate in the synchronization stage of a LoRa receiver are the carrier frequency offset (CFO) and the sampling time offset (STO)~\cite{ghanaatian2019lora,bernier2020low}.
Efficient estimation of these offsets is crucial, as the presence of residual offsets would prevent a receiver to reach the low sensitivity levels offered by the LoRa modulation~\cite{bor2016lora}.
The only studies proposing LoRa receivers robust to CFO are~\cite{ghanaatian2019lora} and~\cite{bernier2020low}, but they do not
entirely address the mitigation of the STO. In~\cite{ghanaatian2019lora}, the STOs are not modeled.
A first low-complexity synchronization algorithm capable of estimating both the STO and the CFO is presented in~\cite{bernier2020low}.
The authors notably show that an STO can be decomposed into integer and fractional parts. Yet, they do not fully examine the
impact of the residual fractional time offset on the demodulation stage and the overall packet error rate (PER).
In their proposed synchronization algorithm, the authors of~\cite{bernier2020low} estimate the fractional STO with a high-variance estimator and correct it
only in the payload of the packets, which strongly degrades the PER. However, we show in this work that it is possible
to obtain significant performance improvements by deriving a new synchronization algorithm based on a more accurate model of the integer and fractional STOs.

The final objective of this paper is to design an efficient synchronization scheme enabling receivers to fully benefit from the advantages of the LoRa modulation.
This requires to design a new estimator for the fractional part of the STO and a new synchronization algorithm that can leverage this estimator.
Specific attention must also be paid to the complexity of the synchronization stage.
Since a LoRa end node may spend more time waiting on a downlink packet than actually demodulating it~\cite{pop2017does}, it is important
that the overall complexity of the synchronization does not dominate the complexity of the demodulation stage.

\subsubsection*{Related work}
The first reverse engineering attempt of the LoRa PHY and a first rough preamble synchronization for LoRa packets is described in~\cite{knight2016decoding},
but no frequency offset estimation and correction is included. The authors in~\cite{robyns2018multi} present a preamble synchronization algorithm for LoRa,
based on the Schmidl-Cox algorithm, that however works only for very high signal-to-noise ratio (SNR) values.
The work in~\cite{ghanaatian2019lora} presents a discussion on the impact of carrier and sampling frequency offsets on LoRa demodulation,
but no discussion is included for the fractional time offset.
The work in~\cite{bernier2020low} presents the first low-complexity synchronization algorithm for LoRa,
including estimators for all integer and fractional time and frequency offsets.
The authors evaluate through simulations the performance of their algorithm using the preamble misdetection rate, but only in the absence of fractional time offsets.
Moreover, no overall packet error rate is evaluated for the proposed algorithm.
The work in~\cite{tapparel2020open} presents a high-complexity SDR implementation of the LoRa PHY in GNU Radio.
The bit error rate of the payload is evaluated using a USRP testbed and shown to be good,
but no evaluation of a preamble misdetection rate or an overall packet error rate is presented.
Recently, the authors in~\cite{savaux2021} present a joint estimator of time and frequency offsets based on the maximum-likelihood principle,
but its very stringent computational complexity do not make it a suitable candidate for low-power IoT end nodes.
Despite the increasing efforts during the last years on the topic of LoRa synchronization and offsets correction,
a low-complexity, yet efficient and complete, LoRa synchronization algorithm that results in low packet error rates for the relative SNRs is still missing in the literature.

\subsubsection*{Contributions} In this work, we build an efficient and low-complexity synchronization algorithm for Nyquist-rate LoRa receivers.
To this end, we derive a complete analytical model of the effects induced by both carrier frequency offsets and sampling time offsets in the demodulation stage.
We notably demonstrate for the first time in the literature that the CFO and STO have distinct effects on the demodulation of LoRa and are not entirely equivalent.
Leveraging this model, we design a new accurate estimator for the fractional STO and show that the estimations of the STO and CFO are actually intertwined.
Since a joint estimation of the CFO and STO would be too complex for low-power end nodes, we design a novel low-complexity synchronization algorithm capable
of precisely estimating and correcting both offsets.
A performance evaluation of the proposed algorithm versus the only other low-complexity algorithm in the literature, namely~\cite{bernier2020low},
underlines the importance of correcting the fractional parts of the CFO and STO before estimating their integer components. Finally, we show that our low-complexity algorithm leads to significantly improved overall packet error rate values that are not more than $1$ or $2$~dB away from the packet error rate of a perfectly synchronized receiver without any offset,
depending on the coding rate.

\subsubsection*{Outline} The remainder of this paper is organized as follows. The working principles of the LoRa PHY are first introduced in Section~\ref{sec:lora}.
The analytical model of a LoRa receiver contaminated by a CFO and STO is obtained in Section~\ref{sec:STO}.
An efficient estimator for the fractional STO is subsequently derived in Section~\ref{sec:estimators}. 
We then design in Section~\ref{sec:receiver} a low-complexity synchronization scheme leveraging the previously obtained results.
Its performance is finally evaluated in Section~\ref{sec:simu}.

\section{Principles of the LoRa PHY}
\label{sec:lora}

The LoRa PHY operates within the unlicensed industrial, scientific and medical (ISM) radio bands, mainly around 868~MHz in Europe and 915~MHz in North America~\cite{adelantado2017understanding}.
The protocol uses a chirp spread spectrum modulation, which brings several benefits for IoT communications. 
Notably, the employed waveform can easily be adapted to trade data throughput for lower energy consumption and for longer communication range~\cite{haxhibeqiri2018survey},~\cite{xu2019measurement}.
This low power consumption enables the deployment of IoT applications where the end nodes are powered only through energy harvesting \cite{sherazi2018energy,sherazi2020energy}.
LoRa is also more robust to frequency selective channels than conventional modulation schemes~\cite{vangelista2017frequency}.
Depending on the scenarios, practical communication ranges vary from 200~m to 30~km~\cite{callebaut2019characterization},~\cite{petajajarvi2015coverage}.

In this section, we present the modulation used in LoRa and the related demodulation methods. The structure of the LoRa preamble is then
detailed since it plays an important role in the synchronization procedure of LoRa receivers.
Finally, the interleaving and coding schemes of the physical layer are introduced.

\subsection{Modulation}

The LoRa modulation relies on chirps, i.e., complex phasors whose instantaneous frequency increases linearly with time over a given bandwidth $B \in \{125, 250, 500\}$ kHz.
Chirps whose frequency increases in time are called \textit{upchirps}, while \textit{downchirps} have an instantaneous frequency that decreases over time.
The duration $T_S$ of a chirp is defined by $T_S = \frac{2^{\mathrm{SF}}}{B}$, where $\mathrm{SF}$ is called the \textit{spreading factor}.
When a receiver samples chirps at the Nyquist frequency $f_S = B$, each chirp contains $N = 2^{\mathrm{SF}}$ samples.
Valid spreading factors as defined by the LoRa Alliance range from 7 to 12 included~\cite{sornin2015lorawan}.
The MAC layer selects the most appropriate spreading factor for each transmission based on the observed wireless conditions~\cite{finnegan2020analysis}.

The symbols are modulated by selecting the initial instantaneous frequency of the chirp, with $N$ possible different initial frequencies. The complex baseband-equivalent representation of
an upchirp $x_s(t)$ modulated with a symbol $s$ between $0$ and $N-1$ is given by~\cite{ghanaatian2019lora}:
\begin{equation} \label{eq:chirp_time}
x_{s}(t)=\left\{\begin{array}{ll}{e^{j 2 \pi\left(\frac{B}{2 T_{s}} t^{2}+B \left(\frac{s}{N}-\frac{1}{2}\right) t\right)}} & {\text{for}~0 \leq t<t_{\mathrm{fold}}} \\ {e^{j 2 \pi\left(\frac{B}{2 T_{s}} t^{2}+B\left(\frac{s}{N}-\frac{3}{2}\right) t\right)}} & {\text{for}~ t_{\mathrm{fold}} \leq t<T_{s}.}\end{array}\right.
\end{equation}

As shown in~\eqref{eq:chirp_time}, the waveforms are built
piecewise around a folding time $t_{fold} = \frac{N - s}{B}$, at which time the instantaneous frequency is decreased by $B$ to keep the signal
in the allocated bandwidth~\cite{seller2016patent}.
Since $N$ different initial frequencies may be selected by the transmitter, this modulation scheme can encode up to $\mathrm{SF}$ information bits 
per chirp.

\subsection{Demodulation}

A receiver perfectly synchronized in time, frequency and phase implements the following steps to demodulate a LoRa symbol.
Let $y_s[n]$ be the sampled signal by the receiver at the Nyquist rate $f_S = B$, where $n \in \{0, \dots, N-1\}$ is the sample index.
The signal $y_s[n]$ is modeled as $y_s[n] = x_s[n] + w[n]$, where 
\begin{equation} \label{eq:chirp_discrete}
    x_s[n] = e^{j 2 \pi \left[\frac{n^2}{2 N} + \left( \frac{s}{N} - \frac{1}{2} \right) n \right]}
\end{equation}
and $w[n]$ is additive white gaussian noise (AWGN)~\cite{chiani2019lora}. Due to the perfect synchronization, the frequency folding
from $\frac{B}{2}$ to $-\frac{B}{2}$ only incurs a phase change of $2\pi$ and can thus be neglected, simplifying the notation of $x_s[n]$.

The receiver processes windows of $N$ samples, each one containing a single symbol.
For every window, the sampled signal $y_s[n]$ is first multiplied point-wise with an unmodulated downchirp $\overline{x}_0[n]$, i.e.,
the complex conjugate of $x_0[n]$. Multiplying the received chirp by $\overline{x}_0[n]$ is called \textit{dechirping}, 
as it removes the squared phase component from $y_s[n]$, but leaves the frequency term that depends on $s$ which carries the modulated information.
We denote the dechirped signal as $\tilde{y}_s[n]$. By dechirping the signal, we obtain
\begin{equation}
    \label{eq:dechirping}
    \begin{aligned}
    \tilde{y}_s[n] &= y_s[n] \cdot \overline{x}_0[n]\\
                &= e^{j 2 \pi \frac{sn}{N}} + \tilde{w}[n],
    \end{aligned}
\end{equation}
with $\tilde{w}[n] = \overline{x}_0[n] \cdot w[n]$. Common demodulation strategies rely on computing the $N$-point discrete Fourier transform (DFT)
of the dechirped signal. We define $Y_k$ and $W_k$ as the result of the DFTs of $\tilde{y}_s[n]$ and $\tilde{w}[n]$, respectively.
Applying the DFT on the dechirped signal yields
\begin{equation}
    Y_k = N \cdot \Pi(k - s, 0) + W_k
\end{equation}
wherein $\Pi(k, \xi) = \dfrac{1 - e^{-j 2 \pi (\xi - k)}}{1 - e^{-j 2 \pi (\xi - k)/N}}$
is the discrete \textit{sinc} function centered around the fractional frequency $\xi$ for an $N$-point DFT.
For a perfectly synchronized receiver we have $\xi = 0$, and the \textit{sinc} function becomes a Kronecker delta. 

The maximum likelihood detector for LoRa symbols in the presence of AWGN consists in computing $Y_k$ 
and selecting the index of the frequency bin with the greatest real part: $\widehat{s} = \argmax_k \real \{ Y_k \}$.
However, this receiver is very sensitive to impairments that induce phase shifts~\cite{marquet2019theo-ber}. 
A non-coherent detection allows to resolve this issue by using the magnitudes of $Y_k$ instead of their real parts: 
\begin{equation} \label{eq:demod}
    \widehat{s} = \argmax_k |Y_k|.
\end{equation}
In the presence of AWGN, the non-coherent detection incurs a $0.7$~dB loss compared to the coherent detection \cite{marquet2019theo-ber,afisiadis2020advantage}.
The non-coherent detector of~\eqref{eq:demod} is the most commonly found throughout the literature~\cite{knight2016decoding,robyns2018multi,marquet2020towards,temim2020enhanced,afisiadis2019error}.

\subsection{Preamble Structure}

\begin{figure}[t]
    \centering
    \includegraphics[width=0.45\textwidth]{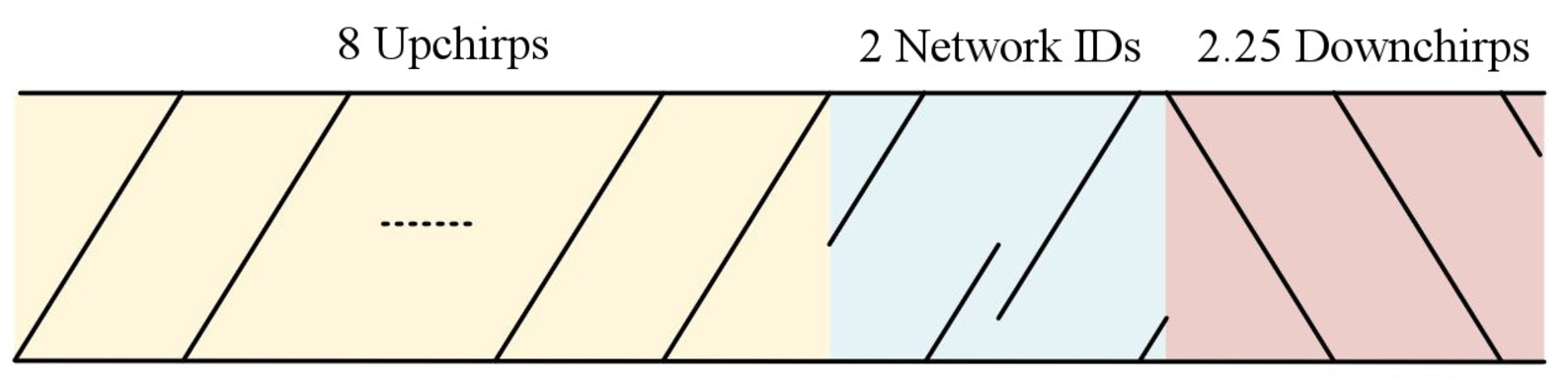}
    \caption{Structure of a LoRa preamble.}
    \label{fig:preamble-struct}
\end{figure}

Every LoRa frame starts with a preamble containing eight repetitions of an unmodulated upchirp $x_0[n]$ ($s = 0$), followed by two
network identifier symbols and two-and-a-quarter repetitions of a downchirp $\overline{x}_0[n]$. An illustration of the LoRa preamble is provided in Fig.~\ref{fig:preamble-struct}.

The structure of the preamble has been designed in order to facilitate the synchronization of the receiver with the transmitter~\cite{seller2016patent}. 
It is explained in~\cite{seller2018patent} and~\cite{bernier2020low} that the joint usage of upchirps and downchirps
in the preamble enables a separate estimation of the CFO and STO.
A downchirp can be demodulated similarly to an upchirp by dechirping the sampled signal with its complex conjugate, i.e., with an unmodulated upchirp $x_0[n]$.

\subsection{Interleaving and Coding}

Beside the modulation and demodulation stages, the physical layer of a LoRa transceiver also includes a whitening scheme, Hamming coding, an interleaving scheme and Gray mapping~\cite{robyns2018multi,afisiadis2020coded}.
LoRa uses $(4,n_c)$ Hamming codes with codeword lengths $n_c \in \{6,7,8\}$.
The conventional $(4,7)$ and extended $(4,8)$ Hamming codes are capable of correcting one bit-error per codeword, whereas the punctured $(4,6)$ Hamming code only allows the detection of two bit errors.
For $n_c = 7$ and $n_c = 8$, it has been shown that the combination of Gray mapping, interleaving, and coding improves the error rates of LoRa receivers subject to residual CFO values \cite{afisiadis2020coded}.

\section{Analytical model of sampling time and carrier frequency offsets}
\label{sec:STO}

In this section, we consider a receiver contaminated by both a CFO $\Delta f_c$ and an STO $\tau$, operating in a noiseless environment.
The continuous-time baseband-equivalent received signal is
\begin{equation} \label{eq:channel}
    y(t) = e^{j 2 \pi t \Delta f_c} x_s(t + \tau).
\end{equation}

\begin{figure}[t]
    \centering
    \setlength\fwidth{0.85\linewidth}
    \setlength\fheight{4cm}
    \includegraphics{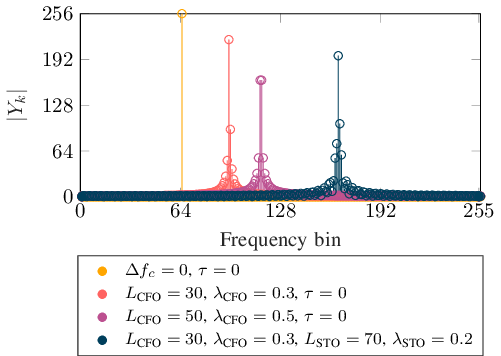}
    \caption{Amplitudes $|Y_k|$ obtained when demodulating a repeating symbol $s = 64$ ($\mathrm{SF} = 8$) without AWGN for different CFO and STO values.
    Integer offsets displace the Kronecker delta originally located at $k = 64$, whereas fractional offsets scatter the delta on several frequency bins.}
    \label{fig:cfo-tmp}
\end{figure}

When demodulating a LoRa symbol $s$, a CFO (resp. STO), displaces the \textit{sinc} function $\Pi(k - s, 0)$ in the frequency domain from position $s$ to
$s + N \Delta f_c / B$ (resp. $s + B\tau$)~\cite{seller2018patent,ghanaatian2019lora,bernier2020low}, as illustrated in Fig.~\ref{fig:cfo-tmp}.
Both offsets may be decomposed into integer and fractional components such as
\begin{equation}
    \label{eq:deltafc-tau}
    \Delta f_c = B \cdot \dfrac{L_{\text{CFO}} + \lambda_{\text{CFO}}}{N}, \tau = \dfrac{L_{\text{STO}} + \lambda_{\text{STO}}}{B}
\end{equation}
with $L_{\text{CFO}}, L_{\text{STO}}$ integer, and $\lambda_{\text{CFO}}, \lambda_{\text{STO}} \in ]-0.5, 0.5]$. Since $L_{\text{CFO}}$ and $L_{\text{STO}}$ are integer,
they shift the Kronecker delta initially at $k = s$ of an integer number of DFT bins.
On the contrary, the fractional offsets $\lambda_{\text{CFO}}$ and $\lambda_{\text{STO}}$ scatter the energy of the symbol
previously contained in a single bin over several frequency bins.

In the presence of AWGN, a symbol error occurs if and only if any of the $|Y_k|$ values for $k \in \{0, \dots, N-1\} \backslash s$ exceeds the value of $|Y_s|$.
We denote the probability of incorrectly demodulating a symbol $s$ under a fractional CFO $\lambda_{\text{CFO}}$ as $P(\hat{s} \neq s | s, \lambda_{\text{CFO}})$.
The scattering induced by the fractional offset increases $P(\hat{s} \neq s | s, \lambda_{\text{CFO}})$,
due to the increase of energy in the bins adjacent to $k = s$.
It is therefore necessary to accurately estimate and correct both integer and fractional offsets to avoid severe degradations of the symbol error rate.

\begin{figure}[t]
    \centering
    \setlength\fwidth{0.85\linewidth}
    \setlength\fheight{5cm}
    \includegraphics{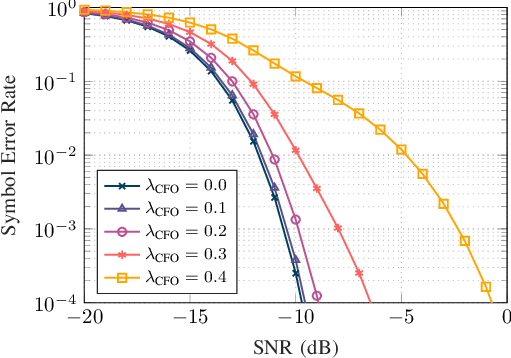}
    \caption{SER of a receiver with a fractional CFO $\lambda_{\text{CFO}}$ ($L_{\text{CFO}} = 0$ and $\mathrm{SF} = 8$) when $\tau = 0$. Identical results apply for a receiver
    with a fractional STO $\lambda_{\text{STO}}$ ($L_{\text{STO}} = 0$) and $\Delta f_c = 0$.}
    \label{fig:ser-cfo}
\end{figure}

An accurate analytical approximation for $P(\hat{s} \neq s | s, \lambda_{\text{CFO}})$ has been derived in~\cite{afisiadis2020coded}.
This approximation is used to illustrate in Fig.~\ref{fig:ser-cfo} the impact of different fractional CFOs $\lambda_{\text{CFO}}$ on the symbol error rate (SER)
for uncoded LoRa.
Obviously, values of $\lambda_{\text{CFO}}$ near $-0.5$ or $0.5$ are much more
detrimental to the SER than smaller ones, e.g., $\lambda_{\text{CFO}} = 0.1$ has almost no impact on the receiver.

Previous works that study synchronization algorithms for Nyquist-rate LoRa receivers do not fully model the impact of the STO on sampled signals.
In~\cite{ghanaatian2019lora}, the integer offsets $L_{\text{CFO}}$ and $L_{\text{STO}}$ are not separated and the fractional time offset $\lambda_{\text{STO}}$ is not included at all in the model.
The authors in~\cite{bernier2020low} include the fractional timing offset $\lambda_{\text{STO}}$ in their discussion and suggest a method to estimate it,
but an exhaustive analytical model of this offset is not examined since they evaluate their proposed synchronization algorithm only under the non-realistic assumption of $\lambda_{\text{STO}}=0$.
In the next subsections, we derive an analytical model for upchirps and downchirps received prior to synchronization, which includes all the aforementioned offsets.
This model then allows us in Section~\ref{sec:estimators} to design a new estimator capable of efficiently estimating the fractional component
of the STO, already during the preamble. Furthermore, in Section~\ref{sec:receiver} we propose a complete synchronization algorithm for LoRa packets, which we evaluate in Section~\ref{sec:simu} in the more realistic scenario of both integer and fractional time and frequency offsets.

\subsection{Continuous-time Model}

We first analytically model the impact of an STO on the receiver, in the concurrent presence of CFO. This requires going back to a continuous-time representation since the discrete-time
representation in~\eqref{eq:chirp_discrete} is only valid for a perfectly synchronized receiver. Let $u(t)$ be the continuous-time step function
and $c(t) = e^{j 2 \pi t \Delta f_c}$ be the added frequency term due to the CFO.
The continuous-time signal $\underline{x}_0(t)$ ($t \in [0, 2 T_S[$) represents two consecutive upchirps in the preamble:
\begin{equation*}
    \underline{x}_{0}(t)= e^{j 2 \pi B \left(\frac{t^{2}}{2 T_{s}} - \frac{t}{2} - t \cdot u(t- T_S) \right)}.
\end{equation*}
Let $y(t)$ be the non-synchronized version of $\underline{x}_{0}(t)$, i.e., after transmission through the channel defined in~\eqref{eq:channel},
where $t \in [0, T_S[$.
Because $y(t)$ is defined over only one symbol period, its expression can be written as the product of an upchirp $x_0(t)$, the effect $c(t)$ of the CFO and a frequency term induced by the STO:
\begin{align*}
    y(t) &= c(t) \underline{x}_0(t + \tau) \\
    &= c(t) e^{j 2 \pi B \big(\frac{t^2 + 2t \tau +\tau^2}{2 T_{s}} - \frac{1}{2} (t + \tau) - (t + \tau) \cdot u(t + \tau - T_S) \big)} \\
    &= x_0(t) c(t) e^{j \left( 2 \pi B \big(\frac{t \tau}{T_{s}} - (t + \tau) \cdot u(t + \tau - T_S) \big) + \theta \right)},
\end{align*}
where $\theta$ is a phase offset representing the constant phase contributions of the STO. This offset does not influence the non-coherent demodulator
\eqref{eq:demod}, as it only uses the magnitudes $|Y_k|$. We hence ignore all constant phase offsets in the following development.

\subsection{Discrete-time Model}
\begin{figure}[t]
    \centering
    \setlength\fwidth{0.85\linewidth}
    \setlength\fheight{3cm}
    \includegraphics{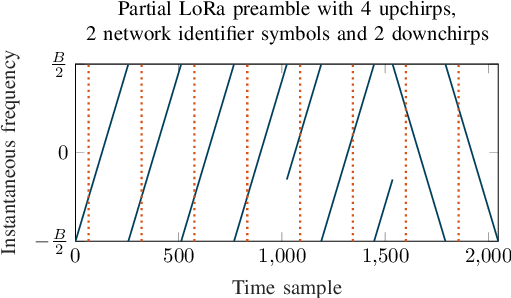}
    \caption{Structure of a preamble for $\mathrm{SF} = 8$. The dotted lines delimitate the consecutive windows of $N$ samples received under 
    an integer STO $L_{\text{STO}} = 64$. Only 4 upchirps are shown for the sake of clarity.}
    \label{fig:preamble}
\end{figure}

The continuous-time signal $y(t)$ is sampled at the frequency $f_s = B$. 
Since a LoRa symbol spans over $N$ samples, a receiver that is not synchronized in time processes windows of samples belonging to two consecutive symbols. For an STO $\tau = (L_{\text{STO}} + \lambda_{\text{STO}})/B$,
the first $M = N - \lfloor L_{\text{STO}} + \lambda_{\text{STO}} \rfloor$ samples in the window belong to the first symbol and the remaining $\lfloor L_{\text{STO}} + \lambda_{\text{STO}} \rfloor$ samples originate
from the second symbol, with $\lfloor \cdot \rfloor$ being the floor operation. This behaviour is illustrated in Fig.~\ref{fig:preamble}.

Since the transmitted signal $\bar{x}_{0}(t)$ consists of two repetitions of the same symbol $x_0(t)$,
it is possible to model analytically the impact of the CFO and STO in the demodulation stage.
Following the definition of $\Delta f_c$ from~\eqref{eq:deltafc-tau}, the discrete-time counterpart of $c(t)$ is
\begin{equation*}
    c[n] = e^{j 2 \pi \frac{L_{\text{CFO}} + \lambda_{\text{CFO}}}{N}n}.
\end{equation*}

Let $y[n]$ be the discrete-time signal sampled from $y(t)$.
This signal is first dechirped by the receiver, i.e., multiplied point-by-point with the sequence $\overline{x}_0[n]$. The expression of the dechirped signal $\tilde{y}[n]$ is:
\begin{align*} 
    \tilde{y}[n] = c[n] e^{j 2 \pi  \Big( \frac{L_{\text{STO}} + \lambda_{\text{STO}}}{N} n 
    - (n + L_{\text{STO}} + \lambda_{\text{STO}}) u[n - M] \Big)},
\end{align*}
where $u[n]$ is the discrete-time step function.
Since $(n + L_{\text{STO}})u[n - M]$ is necessarily integer, this term only induces phases shifts of $2\pi$, and the equation can be simplified to
\begin{equation}
    \label{eq:sto-time-raw}
    \tilde{y}[n] = c[n] e^{j 2 \pi  \left( \frac{L_{\text{STO}} + \lambda_{\text{STO}}}{N} n - \lambda_{\text{STO}} u[n - M] \right)}.
\end{equation}

In~\eqref{eq:sto-time-raw}, we observe that a window of $N$ points sampled with an STO $\tau$ such as $L_{\text{STO}} \neq 0$ and $\lambda_{\text{STO}} \neq 0$
presents an instantaneous phase jump of $-2 \pi \lambda_{\text{STO}}$ at the sample index $M$, i.e., at the boundary of the two unmodulated upchirps.
By separating the fractional offset $\lambda_{\text{STO}}$ from the integer offset $L_{\text{STO}}$, the impact of $\lambda_{\text{STO}}$ on the dechirped signal can be seen as a single frequency component circularly shifted by $M$ samples:
\begin{equation*}
    \tilde{y}[n] = c[n] e^{j 2 \pi \frac{L_{\text{STO}}}{N}n} \cdot \big\langle e^{j 2 \pi \frac{\lambda_{\text{STO}} n}{N}} \big\rangle_{M},
\end{equation*}
where $\langle x[n] \rangle_K$ corresponds to a circular shift of $K$ samples on the signal $x[n]$.

\begin{figure}[t]
    \centering
    \subfloat[Unbounded phase of $\bar{x}_0(t)$.] {
       \setlength\fwidth{0.82\linewidth}
       \setlength\fheight{2.7cm}
       \includegraphics{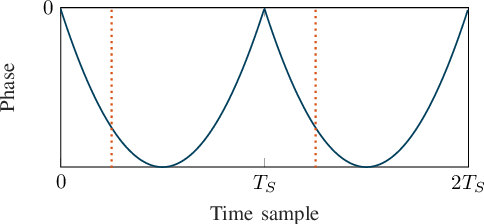}
    }
    
    \subfloat[Phase induced by both offsets in $\tilde{y}{[n]}$.] { 
       \setlength\fwidth{0.82\linewidth}
       \setlength\fheight{2.7cm}
       \includegraphics{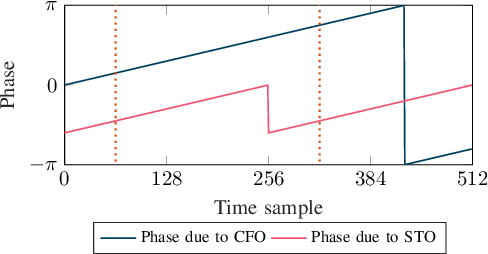}
    }

    \caption{Phases of two consecutive upchirps from the preamble (a) at the transmitter (b) after dechirping, at a receiver contaminated with a CFO $L_{\text{CFO}} = 10$,
    $\lambda_{\text{CFO}} = 0.3$ and an STO $L_{\text{STO}} = 64$, $\lambda_{\text{STO}} = 0.3$. The dotted lines indicate the windows of $N$ samples processed by the receiver.}
    \label{fig:phase-cfo-sto}
\end{figure}

Finally, by expanding the term $c[n]$, we obtain an expression for the dechirped signal depending on all offsets components:
\begin{equation} \label{eq:sto-time}
    \tilde{y}[n] = e^{j 2 \pi \frac{L_{\text{CFO}} + L_{\text{STO}} + \lambda_{\text{CFO}}}{N}n} \cdot \big\langle e^{j 2 \pi \frac{\lambda_{\text{STO}} n}{N}}\big\rangle_{M}.
\end{equation}
Equation~\eqref{eq:sto-time} reveals that
the fractional offsets $\lambda_{\text{CFO}}$ and $\lambda_{\text{STO}}$ cannot be considered equivalent.
Notably, the CFO adds to the dechirped signal $\tilde{y}[n]$ a residual frequency term continuous across the upchirps boundaries, whereas the fractional STO induces a
frequency term whose phase is reset to $-2 \pi \lambda_{\text{STO}}$ after crossing an upchirp boundary.
This difference is reflected in Fig.~\ref{fig:phase-cfo-sto}, where $\lambda_{\text{STO}} = \lambda_{\text{CFO}}$ such that both phase terms increase at the same speed.
The phase induced by the STO is cyclic between the boundaries of symbols, whereas the phase due to the CFO
is continuous across symbols and wraps around $-\pi$ and $\pi$. This difference of behavior is not modeled in~\cite{bernier2020low}, as they consider both fractional offsets to be equivalent.

This absence of equivalence for the fractional offsets leads to a DFT output $Y_k$ that is no longer similar to a \textit{sinc} function in the concurrent presence of both time and frequency offsets. 
Let $a[n] \circledast b[n]$ be the convolution between the signals $a[n]$ and $b[n]$. Computing the $N$-point DFT of~\eqref{eq:sto-time} yields the following signal:
\begin{equation} \label{eq:cfo-sto-freq}
    \begin{split}
    Y_k &= N \cdot \Pi(k - L_{\text{CFO}} - L_{\text{STO}}, \lambda_{\text{CFO}}) \\ 
    & \circledast \Big\lbrack e^{-j 2 \pi \tfrac{k M}{N}} \cdot \Pi(k, \lambda_{\text{STO}}) \Big\rbrack.
    \end{split}
\end{equation}
Due to the term $e^{-j 2 \pi \tfrac{k M}{N}}$ from~\eqref{eq:cfo-sto-freq}, $|Y_k|$ retains a peak around $k = (L_{\text{CFO}} + L_{\text{STO}}) \bmod N$, but its tails do not match the tails of a \textit{sinc} function,
as shown in Fig. \ref{fig:cfo-tmp}.
The expression of $Y_k$ complexifies the estimation of both STO and CFO, as explained in Section~\ref{sec:estimators}.

\subsection{Differences Between Upchirps and Downchirps}

The same development is now used to derive the analytical representation of an unmodulated downchirp contaminated by both offsets.
It can be shown that the expression of a sampled downchirp $y_{d}[n]$ contaminated by an STO $\tau$ and CFO $\Delta f_c$ corresponds to
\begin{align*}
    y_d[n] = c[n] e^{{-} j 2 \pi \left( \frac{n^2}{2 \cdot N} + \frac{L_{\text{STO}}n}{N} - \frac{n}{2} \right)} \cdot \big\langle e^{{-} j 2 \pi \frac{\lambda_{\text{STO}} n}{N}}\big\rangle_{M}.
\end{align*}
After dechirping with an upchirp and expanding $c[n]$, we obtain
\begin{align} \label{eq:cfo-sto-time-downchirp}
    \tilde{y}_d[n] &= e^{j 2 \pi \frac{L_{\text{CFO}} - L_{\text{STO}} + \lambda_{\text{CFO}}}{N} n} \cdot \big\langle e^{{-} j 2 \pi \frac{\lambda_{\text{STO}}}{N} n} \big\rangle_{M}.
\end{align}
We finally denote the DFT of $\tilde{y}_d[n]$ as $Y^d_k$, whose expression is given by
\begin{equation} \label{eq:cfo-sto-freq-downchirp}
    \begin{split}
    Y_k^d &= N \cdot \Pi(k - L_{\text{CFO}} + L_{\text{STO}}, \lambda_{\text{CFO}}) \\
    & \circledast \Big\lbrack e^{-j 2 \pi \tfrac{k M}{N}} \cdot \Pi(k, -\lambda_{\text{STO}}) \Big\rbrack .
    \end{split}
\end{equation}

By comparing~\eqref{eq:cfo-sto-freq} and~\eqref{eq:cfo-sto-freq-downchirp}, we observe that positive STOs and CFOs shift the spectral line in the 
same direction for upchirps, but in opposite directions for downchirps. This property is needed to separately estimate the integer offsets
$L_{\text{CFO}}$ and $L_{\text{STO}}$, as first discussed in~\cite{seller2018patent,bernier2020low} and explained in the following section under our detailed model.

\section{Deriving an accurate estimator for the fractional STO}
\label{sec:estimators}

Due to the complexity of the DFT output $Y_k$ given in~\eqref{eq:cfo-sto-freq}, deriving estimators of the STO and CFO when $L_{\text{CFO}}, L_{\text{STO}}, \lambda_{\text{CFO}}, \lambda_{\text{STO}} \neq 0$ 
is challenging. The difficulty of the problem can however be reduced by correcting partially some offsets.
In this section, we show that estimating $\lambda_{\text{CFO}}$ can be achieved independently of the presence of the other offsets,
but that the estimations of $\lambda_{\text{STO}}$, $L_{\text{STO}}$ and $L_{\text{CFO}}$ are all intertwined.

After correction of the fractional CFO $\lambda_{\text{CFO}}$, the \textit{sinc} function $\Pi(k - L_{\text{CFO}} - L_{\text{STO}}, \lambda_{\text{CFO}})$ from~\eqref{eq:cfo-sto-freq} becomes a Kronecker delta $\delta[n - L_{\text{STO}} - L_{\text{CFO}}]$.
Consequently, the DFT of the partially corrected signal has a simpler expression without any convolution:
\begin{align}
    \label{eq:y-dft-phi-fixed}
    Y_k = N \cdot e^{-j 2 \pi \tfrac{k M}{N}} \cdot \Pi(k - L_{\text{CFO}} - L_{\text{STO}}, \lambda_{\text{STO}}) + W_k.
\end{align}
The energy of the signal in~\eqref{eq:y-dft-phi-fixed} is now spread in multiple DFT bins only due to the impact of $\lambda_{\text{STO}}$. Using~\eqref{eq:y-dft-phi-fixed}, in this section we derive a new efficient estimator for $\lambda_{\text{STO}}$. Moreover, we show that the correction of $\lambda_{\text{STO}}$ before the estimation of $L_{\text{CFO}}$ and $L_{\text{STO}}$ is crucial for the correct estimation of these integer offsets, a fact that has been neglected in the literature~\cite{ghanaatian2019lora,bernier2020low,tapparel2020open}.
To simplify the upcoming equations, we discuss in Section~\ref{sec:lambda_CFO_estim} the estimation and correction of the fractional CFO as a first step performed by the receiver.
All subsequent steps are discussed after this partial correction, thus with $\lambda_{\text{CFO}} = 0$.

\subsection{Estimating $\lambda_{\text{CFO}}$ is Independent of the Other Offsets}
\label{sec:lambda_CFO_estim}

We showed in Section~\ref{sec:STO} that the fractional CFO and STO have distinct effects on the DFT output $Y_k$. The former induces a linear phase term that is continuous across symbols, whereas the phase due to the latter
is cyclic with period $N$. This cyclical property allows a receiver to estimate $\lambda_{\text{CFO}}$ independently of $\lambda_{\text{STO}}$. Let $\tilde{y}^l[n]$ be the $l$-th window of $N$ samples dechirped by the receiver,
and $Y^l_k$ its $N$-point DFT.
Starting from the time-domain representation of unmodulated upchirps from~\eqref{eq:sto-time}, and ignoring the AWGN for now, the received symbols are identical up to a phase difference of $2\pi\lambda_{\text{CFO}}$:
\begin{align*}
    \tilde{y}^{l}[n] &= e^{j 2 \pi \frac{L_{\text{CFO}} +  \lambda_{\text{CFO}}}{N} (n + lN)} \cdot \big\langle e^{j 2 \pi \frac{L_{\text{STO}} + \lambda_{\text{STO}}}{N}n} \big\rangle_{M} \\
    &= e^{j 2 \pi \lambda_{\text{CFO}}} \tilde{y}^{l-1}[n].
\end{align*}
By linearity of the DFT, the same result holds in the frequency domain: $Y^l_k = e^{j 2 \pi \lambda_{\text{CFO}}} Y^{l-1}_k$.
It has been shown that this property can be leveraged to estimate $\lambda_{\text{CFO}}$, by multiplying a DFT bin 
$Y^l_{k}$ with $\overline{Y}^{l-1}_{k}$, the conjugate of the same bin from the previous upchirp, and taking the phase of the product~\cite{bernier2020low}:
\begin{equation} \label{eq:phi-hat}
    \widehat{\lambda}_{\text{CFO}} = \frac{1}{2 \pi} \angleF \left( \sum^{N_{C}}_{l = 2} \sum^{2}_{p = -2} Y^l_{i + p} \cdot \overline{Y}^{l-1}_{i + p} \right),
\end{equation}
with $i = \argmax_k |Y^l_k|$.
To improve the accuracy of the estimate, the five $Y^l_k$ values of each symbol centered around the bin of maximum height are used. In practice,
these DFT bins near $k = (L_{\text{CFO}} + L_{\text{STO}}) \bmod N$ benefit from the processing gain of the modulation, and thus exhibit a higher SNR. 
This operation is repeated over $N_C$ pairs of successive upchirps, and all products $Y^l_{i + p} \cdot \overline{Y}^{l-1}_{i + p}$ are summed up before computing the final estimate.

\subsection{Estimating $L_{\text{CFO}}$ and $L_{\text{STO}}$ Depends on $\lambda_{\text{STO}}$}
\label{subsec:estim-lm}

In Section~\ref{sec:STO}, we showed that an STO has an opposite effect on upchirps and downchirps. 
Following~\eqref{eq:cfo-sto-freq} and~\eqref{eq:cfo-sto-freq-downchirp},
in the absence of AWGN and after correction of  $\lambda_{\text{CFO}}$, the demodulation decision~\eqref{eq:demod} for both an upchirp and a downchirp yields
\begin{align} \label{eq:s_up_down}
    s_{\text{up}} = \argmax_k |Y_k| = (L_{\text{CFO}} + L_{\text{STO}}) \bmod N, \\
    s_{\text{down}} = \argmax_k |Y^d_k| = (L_{\text{CFO}} - L_{\text{STO}}) \bmod N.
\end{align}
By respectively adding and subtracting $s_{\text{up}}$ and $s_{\text{down}}$, $L_{\text{CFO}}$ and $L_{\text{STO}}$ can be estimated separately to some extent.
The estimation of $L_{\text{CFO}}$ is first obtained by adding $s_{\text{up}}$ and $s_{\text{down}}$, and correcting for the modulo effect
from the DFT~\cite{seller2018patent},~\cite{bernier2020low}:
\begin{equation} \label{eq:l-hat}
    \widehat{L}_{\text{CFO}} = \frac{1}{2} \Gamma_{N} \Big[ (s_{\text{up}} + s_{\text{down}}) \bmod N \Big] ,
\end{equation}
where $\Gamma_{N}[k] = \left\{ \begin{array}{ll} k & \text{for~} 0 \leq k < \tfrac{N}{2}, \\
                                             k - N & \text{for~} \tfrac{N}{2} \leq k < N.
                            \end{array} \right. $                 

Once $\widehat{L}_{\text{CFO}}$ is known, $L_{\text{STO}}$ can be estimated using
\begin{equation} \label{eq:m-hat}
     \widehat{L}_{\text{STO}} = (s_{\text{up}} - \widehat{L}_{\text{CFO}}) \bmod N.
\end{equation}
As a consequence of the wrappings induced by the DFT,
it is impossible to recover both integer components in the range $[0, N - 1]$ without ambiguity.
Owing to the fact that the receiver may start acquiring a symbol at any time of its transmission, i.e., $0 \leq L_{\text{STO}} < N$, 
and that $\Delta f_c$ may be positive or negative, the estimation of $L_{\text{CFO}}$ is thus bounded to $[-\tfrac{N}{4}, \tfrac{N}{4} - 1]$~\cite{bernier2020low}.

Since the estimators~\eqref{eq:l-hat} and~\eqref{eq:m-hat} rely on the demodulation stage provided by~\eqref{eq:demod}, the presence of a potential fractional STO
increases the probability of incorrectly demodulating $s_{\text{up}}$ and $s_{\text{down}}$.
To the contrary of the fractional CFO which can be estimated and corrected independently of the other offsets, the estimation of the integer STO and CFO is sensitive to the presence of $\lambda_{\text{STO}}$.
To decrease the probability of a synchronization failure, it is thus crucial to estimate and correct the fractional STO, even partially, before estimating the integer offsets.
We note that in the literature~\cite{ghanaatian2019lora,bernier2020low,tapparel2020open}, the fractional STO $\lambda_{\text{STO}}$ is never corrected before the integer offsets,
which strongly increases the probability of wrongly estimating the integer offsets and thus of experiencing a synchronization failure.
None of the aforementioned works shows the importance of correcting $\lambda_{\text{STO}}$ before the integer offsets,
since none of them presents results for the overall packet error rate in the presence of all integer and fractional offsets.
In particular, \cite{tapparel2020open} shows good conditional bit error rates (i.e., only for the packets that managed to synchronize correctly),
but the authors do not indicate how many of the overall transmitted packets actually managed to synchronize.
Moreover, the works in~\cite{ghanaatian2019lora,bernier2020low} simulate results only under the assumption of $\lambda_{\text{STO}}=0$.

\subsection{Estimating the Fractional STO $\lambda_{\text{STO}}$}
\label{subsec:estim-lambda}

Many estimators in the literature have been designed to estimate a fractional frequency $\xi$ of a \textit{sinc} signal $\Pi(k, \xi)$ with AWGN
(e.g.,~\cite{quinn1997estimation,jacobsen2007fast,yang2011noniterative}).
However, due to the additional phase term $e^{-j 2 \pi \tfrac{k M}{N}}$ in~\eqref{eq:y-dft-phi-fixed}, the expression of the signal $Y_k$ after correction of
$\lambda_{\text{CFO}}$ is not a strict \textit{sinc} function.
Estimating $\lambda_{\text{STO}}$ from usual estimators therefore requires \textit{a priori} knowledge of
$M = N - \lfloor L_{\text{STO}} + \lambda_{\text{STO}} \rfloor$, which is unknown until the reception of a downchirp.

Yet, as previously discussed, the fractional STO should preferably be corrected before estimating $L_{\text{STO}}$.
To avoid the complex joint estimation of $L_{\text{STO}}$ and $\lambda_{\text{STO}}$ over several symbols, we suggest
instead to design an estimator $\widehat{\lambda}_{\text{STO}}$ that relies on a prior estimate $\widehat{M}$.
This estimator $\widehat{\lambda}_{\text{STO}}$ will then be used iteratively in the synchronization algorithm, with
increasingly accurate estimates of $M$.

Regarding first the specific case when $M = 0$ (i.e., when $Y_k$ is a \textit{sinc} function with AWGN), the following estimator from~\cite{jacobsen2007fast} is valid:
\begin{equation} \label{eq:estimator-nozp}
    \widehat{\lambda}_{\text{STO}} = -\real \Big\lbrack \dfrac{Y_{i+1} - Y_{i-1}}{2 Y_i - Y_{i+1} - Y_{i-1}} \Big\rbrack,
\end{equation}
with $i = \argmax_k |Y_k|$.
We now extend~\eqref{eq:estimator-nozp} for the general case when the receiver is affected by an STO so that $M \neq 0$.
This requires correcting the phases of $Y_k$ to remove the term $e^{-j 2 \pi \tfrac{k M}{N}}$. 
Assuming the knowledge of an estimation of $L_{\text{STO}}$, and using the approximation $\widehat{M} \approx N - L_{\text{STO}}$ to correct the phases of $Y_k$, we obtain an estimator
relying on the complex values $Y_k$ of a demodulated upchirp:
\begin{equation} \label{eq:estimator-lambda}
    \widehat{\lambda}_{\text{STO}} = -\real \left\lbrack \dfrac{e^{-j 2 \pi \tfrac{\widehat{M}}{N}} Y_{i+1} - e^{j 2 \pi \tfrac{\widehat{M}}{N}} Y_{i-1}}
                        {2 Y_i - e^{-j 2 \pi \tfrac{\widehat{M}}{N}} Y_{i+1} - e^{j 2 \pi \tfrac{\widehat{M}}{N}} Y_{i-1}} \right\rbrack .
\end{equation}
The precision of this estimator depends on the accuracy of $\widehat{M}$ itself.
The accuracy of the estimation can also be improved by averaging the complex values $\{Y_{i-1}, Y_i, Y_{i+1}\}$ over several upchirps.

To start estimating the fractional STO without an estimate $\widehat{M}$ and thus without having to wait on a downchirp, another estimator $\widehat{\lambda}_{\overline{\text{STO}}}$
that relies only on the magnitudes $|Y_k|$ is proposed in~\cite{seller2018patent} and used in~\cite{bernier2020low}. This estimator leverages the intermediate function
\begin{equation*}
    T(\lambda_{\text{STO}}) = \dfrac{|\Pi(1, \lambda_{\text{STO}})| - |\Pi(-1, \lambda_{\text{STO}})|}{|\Pi(0, \lambda_{\text{STO}})|}.
\end{equation*}
After demodulating an upchirp, the receiver estimates $\lambda_{\text{STO}}$ by inverting $T(\lambda_{\text{STO}})$ as follows:
\begin{equation} \label{eq:lambda-hat-bernier}
    \widehat{\lambda}_{\overline{\text{STO}}} = T^{-1} \left(\dfrac{|Y_{i+1}| - |Y_{i-1}|}{|Y_i|} \right),
\end{equation}
with $i = \argmax_k |Y_k|$. Averaging the magnitudes $|Y_k|$ of consecutive upchirps enhances the accuracy of~\eqref{eq:lambda-hat-bernier},
a fact that is however not evaluated though in the aforementioned works.
Although $\widehat{\lambda}_{\overline{\text{STO}}}$ is independent of $L_{\text{STO}}$,
this estimator presents a very high variance due to the derivative of $T^{-1}(x)$ being close to 0 for small values of $\lambda_{\text{STO}}$~\cite{seller2018patent}.
We show in Section~\ref{sec:simu} that the high variance of this estimator degrades the receiver performance when used naively in a symbol-by-symbol tracking loop as proposed in~\cite{bernier2020low}.
\begin{figure}[t]
    \centering
    \setlength\fwidth{0.75\linewidth}
    \setlength\fheight{6cm}
    \includegraphics{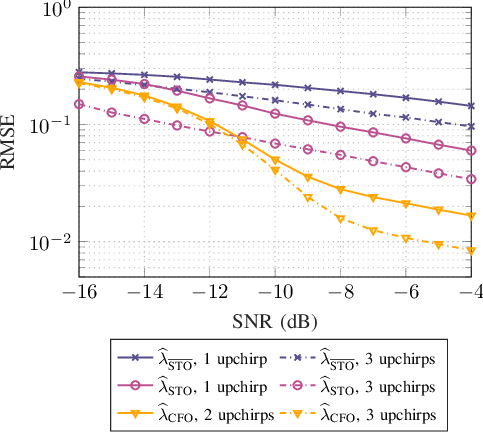}
    \caption{RMSE of the fractional offsets estimators, with $\mathrm{SF} = 8$ and one or several upchirps.}
    \label{fig:rmse}
\end{figure}

The Root Mean Square Errors (RMSE) of all three estimators~\eqref{eq:phi-hat},~\eqref{eq:lambda-hat-bernier} and~\eqref{eq:estimator-lambda} have been
obtained through simulation for $\mathrm{SF} = 8$ and are presented in Fig.~\ref{fig:rmse}. The estimators $\widehat{\lambda}_{\text{STO}}$ and $\widehat{\lambda}_{\overline{\text{STO}}}$ are evaluated
using upchirps contaminated by an STO $\tau$ uniformly distributed in the range $\mathclose[0, \tfrac{N}{B}\mathopen[$. For $\widehat{\lambda}_{\text{STO}}$, it is assumed that the receiver has ideal knowledge of the integer offset $L_{\text{STO}}$.
To evaluate the RMSE of $\widehat{\lambda}_{\text{CFO}}$ from~\eqref{eq:phi-hat}, a fractional CFO $\lambda_{\text{CFO}}$ uniformly distributed in the range $\mathopen]-0.5, 0.5\mathclose]$ is added.
The estimators are subsequently evaluated with two or three upchirps. The averaging of these upchirps is performed differently depending on the estimator, as previously described for each estimator.

Using a single upchirp and for an SNR of $-9$~dB, corresponding to a SER of $10^{-4}$ for a perfectly synchronized receiver, our estimator $\widehat{\lambda}_{\text{STO}}$ is one order of magnitude more accurate than $\widehat{\lambda}_{\overline{\text{STO}}}$ from~\cite{seller2018patent}.
Moreover, using three upchirps instead of one improves the precision of the estimates by a factor $1.9$ for $\widehat{\lambda}_{\text{STO}}$,
but only by a factor $1.3$ for $\widehat{\lambda}_{\overline{\text{STO}}}$.

\section{Design of an efficient low-complexity synchronization algorithm}
\label{sec:receiver}

\begin{figure*}[t]
    \centering
    \includegraphics[width=\textwidth]{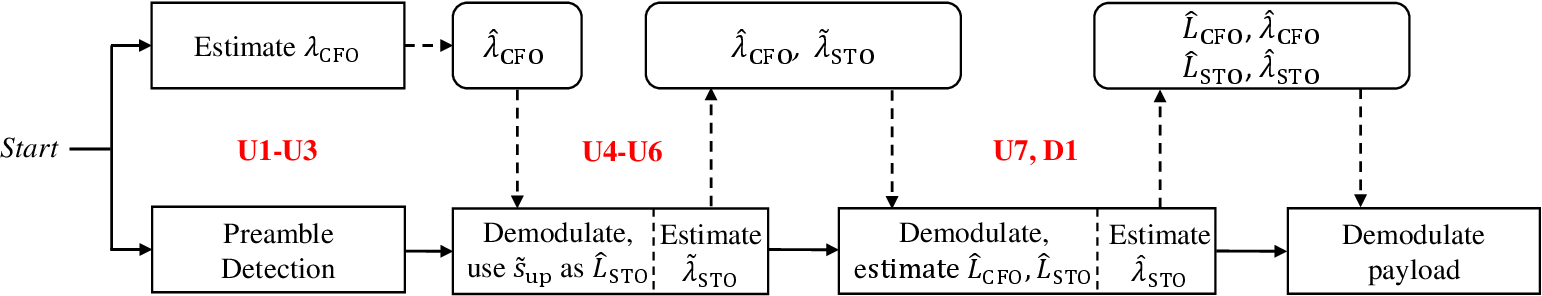}
    \caption{Flowgraph of the synchronization algorithm. For each step, the related parts of the preamble are indicated in red. U$n$ refers to the $n$-th full upchirp in the preamble, and D1 to the single full downchirp.
    Dashed arrows indicate the CFO and STO components estimated or corrected by each stage.}
    \label{fig:flowgraph}
\end{figure*}

We showed in the previous section that the estimation of the offsets components $L_{\text{STO}}$, $L_{\text{CFO}}$ and $\lambda_{\text{STO}}$ are intertwined.
Only the estimation of the fractional CFO $\lambda_{\text{CFO}}$ can be performed independently of the other offsets.
Moreover, since the estimation of the integer offsets requires the demodulation of a downchirp, which is only available at the end of the preamble, 
the interdependencies between offsets complexify the design of a synchronization scheme.
To solve this issue while keeping the implementation of the receiver as simple as possible, we present
in this section a low-complexity synchronization algorithm capable of accurately estimating all offsets components.
The proposed synchronization algorithm is then compared to the synchronization algorithm from~\cite{bernier2020low}.
Both algorithms work on a non-synchronized signal sampled at the Nyquist-rate $f_S = B$ that follows the model described in Section~\ref{sec:STO}.

\subsection{Proposed Synchronization Algorithm}

The proposed algorithm, whose pseudocode is given in Algorithm~\ref{algo:proposed}, leverages sequentially the different parts of the preamble, by processing a single symbol at a time.
It uses the estimators introduced in Section~\ref{sec:estimators}, and estimates the fractional STO $\lambda_{\text{STO}}$ iteratively in two steps.
The algorithm, is split in three stages, as shown in Fig.~\ref{fig:flowgraph}: first the estimation of $\lambda_{\text{CFO}}$, then a preliminary estimation of the fractional STO,
and finally the definitive estimation of $L_{\text{CFO}}$, $L_{\text{STO}}$ and $\lambda_{\text{STO}}$.

\subsubsection{Preamble Detection and Estimation of the Fractional CFO}
Before going into the synchronization algorithm, the receiver first implements a preamble detection scheme.
It is well-known that the presence of several consecutive upchirps can be used to detect the start of a frame \cite{ghanaatian2019lora,robyns2018multi}.
A preamble is detected by a receiver when the same symbol $s$, or its neighbors $s \pm 1$, are consecutively demodulated~\cite{tapparel2020open}.
Let $y^l[n]$ be the $l$-th window of $N$ samples processed by the receiver.
Upon a preamble detection on three consecutive upchirps $y^l[n]$, $y^{l-1}[n]$ and $y^{l-2}[n]$, these symbols are also used to compute the estimate $\widehat{\lambda}_{\text{CFO}}$.
This estimate is used to correct the fractional CFO when demodulating the remaining symbols of the preamble.

\subsubsection{First Correction of the Fractional STO}
The next three upchirps of the preamble are dedicated to the estimation of the fractional STO.
The demodulated value $\tilde{s}_{\text{up}}$ of the fourth upchirp of the preamble is used as an initial estimation of $s_{\text{up}}$.
A temporary estimate $\tilde{\lambda}_{\text{STO}}$ of the fractional STO is then computed using the estimator $\widehat{\lambda}_{\text{STO}}$ from~\eqref{eq:estimator-lambda} with
the averaged DFTs of the three upchirps and $\tilde{s}_{\text{up}}$ as an approximation of $L_{\text{STO}}$, i.e., $\widetilde{M} =  N - \tilde{s}_{\text{up}}$.
The receiver subsequently realigns itself in time to correct the fractional STO using $\tilde{\lambda}_{\text{STO}}$.
This first correction of the fractional STO increases the probability of correctly demodulating the symbols $s_{\text{up}}$ and $s_{\text{down}}$
in the final stage of the algorithm.

Since the estimate $\widetilde{M}$ used in the computation of $\tilde{\lambda}_{\text{STO}}$ is
likely to be close to $N - L_{\text{CFO}} - L_{\text{STO}}$ instead of $N - L_{\text{STO}}$, the accuracy of the estimate $\tilde{\lambda}_{\text{STO}}$ depends on $L_{\text{CFO}}$.
We notably show in Section~\ref{sec:simu} that large offset values $L_{\text{CFO}}$ decrease the precision of $\tilde{\lambda}_{\text{STO}}$,
and hence also the probability of correctly estimating the integer offsets.
Finally, as the definitive estimation of $\widehat{\lambda}_{\text{STO}}$ using~\eqref{eq:estimator-lambda} can only take place once $L_{\text{STO}}$ is estimated,
the averaged DFT bins are stored in memory until the reception of a downchirp.

\subsubsection{Final Synchronization Stage}

The final entire upchirp and the single entire downchirp are used to estimate $\hat{s}_{\text{up}}$ and $\hat{s}_{\text{down}}$.
With these values, the receiver then uses~\eqref{eq:l-hat} and~\eqref{eq:m-hat} to estimate $L_{\text{CFO}}$ and $L_{\text{STO}}$,
respectively.
A definitive estimation of $\lambda_{\text{STO}}$ is afterwards computed using~\eqref{eq:estimator-lambda}, with $\widehat{M} = N - \widehat{L}_{\text{STO}}$ and the DFT bins stored in memory.
The receiver performs the final correction of the CFO and STO using the above estimates and proceeds to the demodulation of the payload.

\subsubsection{Processing of the Payload}
After the final synchronization stage, the receiver proceeds to the demodulation of the $N_P$ payload symbols as explained in Section~\ref{sec:lora}.

\begin{algorithm}[t]
    \caption{Proposed synchronization algorithm}
    \label{algo:proposed}
    \begin{algorithmic}[1]
    \STATE $l = 0$
    \\ \textit{Preamble Detection and Fract. CFO Estimation:}
     \WHILE {preamble not detected}
     \STATE $l = l + 1$
     \STATE $Y^l_k = \text{DFT} \left( y^l[n] \odot \overline{x}_0[n] \right)$, $\hat{s}^l = \argmax_k \left| Y^l_k \right|$
     \STATE $z^l = \sum^{2}_{p = -2} Y^l_{\hat{s}^l + p} \cdot \overline{Y}^{l-1}_{\hat{s}^l + p}$
     \STATE $\text{PreambleDetection}\left( \hat{s}^l, \hat{s}^{l-1}, \hat{s}^{l-2} \right)$
     \ENDWHILE
     \STATE $\widehat{\lambda}_{\text{CFO}} = \frac{1}{2 \pi} \angleF \left(\sum^{1}_{i = 0} z^{l-i} \right)$
     \\\hrulefill
     \\ \textit{First Correction of the Fractional STO:}
     \FOR {$m = 1$ to $3$}
     \STATE $Y^{l+m}_k = \text{DFT} \left( y^{l+m}[n] \odot \overline{x}_0[n] \odot e^{-j 2 \pi \widehat{\lambda}_{\text{CFO}} \frac{mN + n}{N}} \right)$
     \ENDFOR
     \STATE $\tilde{s}_{\text{up}} = \argmax_k |Y^{l+1}_k|$, $\widetilde{M} =  N - \tilde{s}_{\text{up}}$
     \STATE $Y^{\text{avg}}_k = \sum^3_{m=1} Y^{l+m}_k$
     \STATE Compute $\tilde{\lambda}_{\text{STO}}$ with $\{Y^{\text{avg}}_{\tilde{s}_{\text{up}}-1}, Y^{\text{avg}}_{\tilde{s}_{\text{up}}}, Y^{\text{avg}}_{\tilde{s}_{\text{up}}+1}\}$, $\widetilde{M}$ using~\eqref{eq:estimator-lambda}
     \STATE Realign receiver by $\tilde{\lambda}_{\text{STO}}$ samples
     \\\hrulefill
     \\ \textit{Final Synchronization Stage:}
     \STATE $Y^{l+4}_k = \text{DFT} \left( y^{l+4}[n] \odot \overline{x}_0[n] \odot e^{-j 2 \pi \widehat{\lambda}_{\text{CFO}} \frac{n}{N}} \right)$
     \STATE $Y^{l+8}_k = \text{DFT} \left( y^{l+8}[n] \odot x_0[n] \odot e^{-j 2 \pi \widehat{\lambda}_{\text{CFO}} \frac{n}{N}} \right)$
     \STATE $\hat{s}_{\text{up}} = \argmax_k \left| Y^{l+4}_k \right|$, $\hat{s}_{\text{down}} = \argmax_k \left| Y^{l+8}_k \right|$
     \STATE Compute $\widehat{L}_{\text{CFO}}$, $\widehat{L}_{\text{STO}}$ with $\hat{s}_{\text{up}}$, $\hat{s}_{\text{down}}$ using~\eqref{eq:l-hat} and~\eqref{eq:m-hat}
     \STATE Compute $\widehat{\lambda}_{\text{STO}}$ with $\{Y^{\text{avg}}_{\hat{s}_{\text{up}}-1}, Y^{\text{avg}}_{\hat{s}_{\text{up}}}, Y^{\text{avg}}_{\hat{s}_{\text{up}}+1}\}$ and ${\widehat{M} = N - \widehat{L}_{\text{STO}}}$ using~\eqref{eq:estimator-lambda}
     \STATE Correct carrier frequency by $B/N \left( \widehat{L}_{\text{CFO}} + \widehat{\lambda}_{\text{CFO}} \right)$ Hz
     \STATE Realign receiver by $\widehat{L}_{\text{STO}} + \widehat{\lambda}_{\text{STO}} - \tilde{\lambda}_{\text{STO}} + N/4$ samples
     \\\hrulefill
     \\ \textit{Processing of the Payload:}
     \FOR {$m = l+5$ to $l+N_P+4$}
     \STATE $Y^{m}_k = \text{DFT} \left( y^{m}[n] \odot \overline{x}_0[n] \right)$, $\hat{s}_m = \argmax_k \left| Y^{m}_k \right|$
     \ENDFOR
    \end{algorithmic} 
\end{algorithm}

\subsection{Synchronization Algorithm from~\cite{bernier2020low}}

\begin{algorithm}[t]
    \caption{Synchronization algorithm from~\cite{bernier2020low}}
    \label{algo:bernier}
    \begin{algorithmic}[1]
    \STATE $l = 0$
    \\ \textit{Preamble Detection and Fract. CFO Estimation:}
     \WHILE {preamble not detected}
     \STATE $l = l + 1$
     \STATE $Y^l_k = \text{DFT} \left( y^l[n] \odot \overline{x}_0[n] \right)$, $\hat{s}^l = \argmax_k \left| Y^l_k \right|$
     \STATE $z^l = \sum^{2}_{p = -2} Y^l_{\hat{s}^l + p} \cdot \overline{Y}^{l-1}_{\hat{s}^l + p}$
     \STATE $\text{PreambleDetection}\left( \hat{s}^l, \hat{s}^{l-1}, \dots, \hat{s}^{l-5} \right)$
     \ENDWHILE
     \STATE $\widehat{\lambda}_{\text{CFO}} = \frac{1}{2 \pi} \angleF \left(\sum^{4}_{i = 0} z^{l-i} \right)$
     \\\hrulefill
     \\ \textit{Coarse Synchronization Stage:}
     \STATE $Y^{l+1}_k = \text{DFT} \left( y^{l+1}[n] \odot \overline{x}_0[n] \odot e^{-j 2 \pi \widehat{\lambda}_{\text{CFO}} \frac{n}{N}} \right)$
     \STATE $Y^{l+5}_k = \text{DFT} \left( y^{l+5}[n] \odot x_0[n] \odot e^{-j 2 \pi \widehat{\lambda}_{\text{CFO}} \frac{n}{N}} \right)$
     \STATE $\hat{s}_{\text{up}} = \argmax_k \left| Y^{l+1}_k \right|$, $\hat{s}_{\text{down}} = \argmax_k \left| Y^{l+5}_k \right|$
     \STATE Compute $\widehat{L}_{\text{CFO}}$, $\widehat{L}_{\text{STO}}$ with $\hat{s}_{\text{up}}$, $\hat{s}_{\text{down}}$ using~\eqref{eq:l-hat} and~\eqref{eq:m-hat}
     \STATE Correct carrier frequency by $B/N \left( \widehat{L}_{\text{CFO}} + \widehat{\lambda}_{\text{CFO}} \right)$ Hz
     \STATE Realign receiver by $\widehat{L}_{\text{STO}} + N/4$ samples
     \STATE Compute $\widehat{\lambda}^{l+5}_{\overline{\text{STO}}}$ with $\{Y^{l+5}_{\hat{s}_{\text{down}}+1}, Y^{l+5}_{\hat{s}_{\text{down}}}, Y^{l+5}_{\hat{s}_{\text{down}}-1}\}$ using~\eqref{eq:lambda-hat-bernier}
     \\\hrulefill
     \\ \textit{Processing of the Payload:}
     \FOR {$m = l+6$ to $l+N_P+5$}
     \STATE $\Lambda^{m}_{\text{STO}} = \frac{1}{m-l-5}\sum_{k=l+5}^{m-1} \Lambda^{k}_{\text{STO}} + \widehat{\lambda}^{k}_{\overline{\text{STO}}}$
     \STATE Realign receiver by $\Lambda^{m}_{\text{STO}}  - \Lambda^{m-1}_{\text{STO}} $ samples
     \STATE $Y^{m}_k = \text{DFT} \left( y^{m}[n] \odot \overline{x}_0[n] \right)$, $\hat{s}_m = \argmax_k \left| Y^{m}_k \right|$
     \STATE Compute $\widehat{\lambda}^m_{\overline{\text{STO}}}$ with $\{Y^{m}_{\hat{s}_m-1}, Y^{m}_{\hat{s}_m}, Y^{m}_{\hat{s}_m+1}\}$ using~\eqref{eq:lambda-hat-bernier}
     \ENDFOR
    \end{algorithmic} 
\end{algorithm}

We now briefly introduce the synchronization algorithm presented in~\cite{bernier2020low}, whose pseudocode is given in Algorithm~\ref{algo:bernier} to ease the comparison.
In~\cite{bernier2020low}, the preamble detection and the estimation of the fractional CFO are performed using all the upchirps of the preamble, except the last one.
The fractional CFO is then corrected for the remainder of the preamble.
The final upchirp and one single entire downchirp are used to obtain the symbols $\hat{s}_{\text{up}}$ and $\hat{s}_{\text{down}}$, respectively.
The receiver subsequently estimates $L_{\text{CFO}}$ and $L_{\text{STO}}$ using $\hat{s}_{\text{up}}$ and $\hat{s}_{\text{down}}$, and performs a coarse synchronization
which corrects the carrier frequency with $\widehat{L}_{\text{CFO}} + \widehat{\lambda}_{\text{CFO}}$ and the integer STO using $\widehat{L}_{\text{STO}}$, but not the fractional STO.

The fractional STO is estimated using a tracking loop that relies on the estimator $\widehat{\lambda}_{\overline{\text{STO}}}$ of~\eqref{eq:lambda-hat-bernier}.
This tracking loop is enabled after the demodulation of the downchirp, i.e., during the payload, and performs a symbol-by-symbol estimation of $\lambda_{\text{STO}}$.
Since the STO is considered constant, its correction for the $m$-th symbol
can be improved by averaging all the previous estimates of the fractional STO.
This averaging is not discussed in~\cite{bernier2020low}, however we include it in the algorithm for a fair comparison, since it improves the performance of their receiver.

Unlike the proposed algorithm, the algorithm from~\cite{bernier2020low} does not leverage the upchirps of the preamble to estimate and correct the fractional STO
before the estimation of the integer offsets.
As a consequence, the presence of the uncorrected $\lambda_{\text{STO}}$ increases the probability of a wrong demodulation of the symbols $\hat{s}_{\text{up}}$ and $\hat{s}_{\text{down}}$.
Since these symbols are used to compute $L_{\text{STO}}$ and $L_{\text{CFO}}$,
erroneous estimations of the integer offsets frequently arise and strongly deteriorate the performance of the receiver, as shown in Section~\ref{sec:simu}.
We also show next that the usage of the high-variance estimator from~\eqref{eq:lambda-hat-bernier} in a symbol-by-symbol tracking loop further degrades the performance of the receiver.

\section{Performance evaluation}
\label{sec:simu}

In this section, we assess the performance of our proposed synchronization algorithm and we compare it to the performance of the only other low-complexity algorithm in the literature,
i.e., the algorithm from~\cite{bernier2020low}. We first compare the two algorithms under the simulation parameters used in~\cite{bernier2020low}
and we show that they have similar performance.
We then compare the two algorithms under more realistic simulation parameters and show that our algorithm results in a significant improvement of the packet error rate
compared to~\cite{bernier2020low}.
In addition, we analyze the impact of large CFO values on the proposed algorithm. Finally, the error rates of both receivers for coded LoRa packets are presented and compared to the performance of a theoretical perfectly synchronized receiver.

\subsection{Simulation Parameters}

For both our synchronization algorithm and the algorithm in~\cite{bernier2020low}, the receiver samples a signal of bandwidth $B = 125$ kHz with AWGN at a carrier frequency of $f_c = 868$ MHz.
The bandwidth of the signal does not impact the performance of the receiver, except for the range of the CFO, as explained below.
All error rates are evaluated through Monte-Carlo simulations with $10^5$ trials per SNR level.
Each packet starts with a preamble, as described in Section~\ref{sec:lora}. The payload consists of $N_P = 28$ modulated LoRa symbols, either coded or uncoded.
The packets reach the receiver through the channel defined in~\eqref{eq:channel}.
The preamble detection schemes are considered ideal for both algorithms, i.e., the preamble of a packet is always detected, and do not
impact the simulation results.

To compare the two algorithms in a realistic simulation setup, we allow the sampling time offset to take values uniformly in the range $\tau \in \left[0, \tfrac{N}{B}\right[$, except for Fig.~\ref{fig:frame-error-bernier}, where we follow the simulation setup used in~\cite{bernier2020low}.
In particular, for Fig.~\ref{fig:frame-error-bernier}, we allow the STO $\tau$ to take only integer values, uniformly chosen from the set $L_{\text{STO}} \in \{0, \dots, N-1\}$, and we set the fractional time offset $\lambda_{\text{STO}}=0$, exactly as in~\cite{bernier2020low}.
The receivers sample the signal at an oversampled frequency $f'_S = RB$, with $R = 10$, to allow a correction of the fractional time offset.
For both algorithms, the oversampled signal is decimated by $R$ before the dechirping stage, i.e., synchronization and demodulation are done at the Nyquist rate.
When a receiver realigns itself in time, it selects the decimation index among the $R$ available that is the closest to the fractional STO estimated by the synchronization algorithm.

\begin{table}[t]
    \centering
        \begin{tabular}{l|*{3}r}
        \toprule
        \diagbox{$f_c$}{$B$} & $125$kHz & $250$kHz & $500$kHz \\
        \midrule
        $433$~MHz & $\pm 0.07N$ & $\pm 0.03N$ & $\pm 0.02N$ \\
        $868$~MHz & $\pm 0.14N$ & $\pm 0.07N$ & $\pm 0.03N$ \\
        \bottomrule
        \end{tabular}%
      
    \caption{Maximum $L_{\text{CFO}}$ offset observed by a LoRa receiver with a crystal precision of 20 ppm for the different standardized bandwidths and ISM carrier frequencies.}
    \label{tab:max-cfo}
\end{table}

As explained in Section~\ref{sec:estimators} and similarly to~\cite{bernier2020low}, the LoRa receiver studied in this work can only estimate a CFO $\Delta f_c \in \left[-\tfrac{B}{4}, \tfrac{B}{4}\right]$.
Considering a typical carrier frequency of $f_c = 868$~MHz and the smallest
bandwidth $B = 125$~kHz, only CFOs $\Delta f_c$ up to 36 ppm can be estimated. 
Low-power IoT end nodes are usually designed with a low-cost bill of materials, and the crystal oscillators used for the carrier frequency synthesis often have limited accuracy.
For instance, the Semtech SX1276 receiver embeds a crystal with a 20 ppm precision~\cite{semtech:sx1276}, whereas gateways use more accurate temperature-controlled oscillators.
Table~\ref{tab:max-cfo} presents the maximum $L_{\text{CFO}}$ observable among all possible carrier frequencies and bandwidths for a crystal precision of 20 ppm.
The chosen simulation parameters $B = 125$~kHz and $f_c = 868$~MHz are hence the ones inducing the largest CFOs.
In our simulations, the CFO is always uniformly distributed in the range $\Delta f_c \in [-20, 20]$ ppm~\cite{semtech:sx1276}, unless if specified otherwise.
The estimated CFO is corrected by shifting in frequency the sampled signal, as explained for each algorithm in Section~\ref{sec:receiver}.

\subsection{Performance under the Simulation Parameters used in~\cite{bernier2020low}}

\begin{figure}[t]
    \centering
    \setlength\fwidth{0.8\linewidth}
    \setlength\fheight{6.22cm}
    \includegraphics{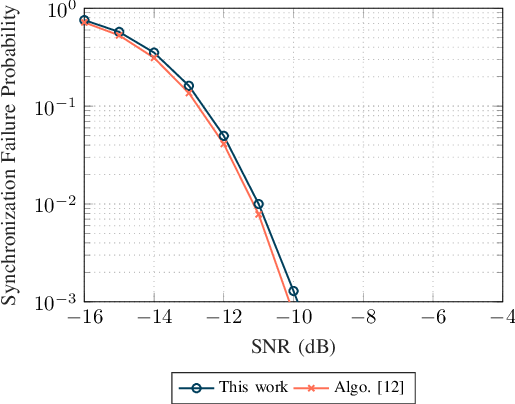}
    \caption{Packet synchronization failure probability following the definition of~\cite{bernier2020low}
    with $\lambda_{\text{STO}} = 0$ and $\textrm{SF} = 8$.}
    \label{fig:frame-error-bernier}
\end{figure}

We first show in Fig.~\ref{fig:frame-error-bernier} the packet synchronization failure probability, as defined in~\cite{bernier2020low},
of both receivers for a spreading factor $\text{SF}=8$.
In~\cite{bernier2020low}, the authors do not simulate a fractional STO, i.e., $\lambda_{\text{STO}}=0$, and a packet synchronization failure
is defined as the event that the algorithm was unable to identify the first data sample, hence leading to a complete packet loss.
We note here that this definition of a synchronization failure from~\cite{bernier2020low} essentially corresponds to the presence of a residual integer time offset
after the synchronization stage, i.e., $\widehat{L}_{\text{STO}} \neq L_{\text{STO}}$.
Exactly as in Fig.~11 of~\cite{bernier2020low}, the results presented in Fig.~\ref{fig:frame-error-bernier} are obtained by setting the fractional STO to zero
and allowing the integer STO to be uniformly chosen from the set $L_{\text{STO}} \in \{0, \dots, N-1\}$.
We observe that under the chosen simulation parameters ($\lambda_{\text{STO}}=0$), both our proposed algorithm and the algorithm from~\cite{bernier2020low}
exhibit almost identical synchronization failure probabilities.

\subsection{Performance under More Realistic Simulation Parameters}

\begin{figure}[t]
	\centering
	\setlength\fwidth{0.8\linewidth}
	\setlength\fheight{10cm}
    \includegraphics{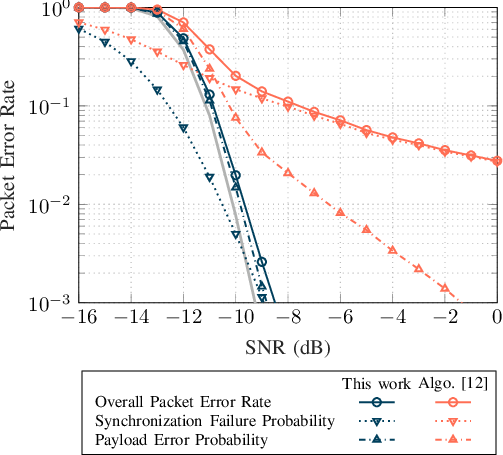}
	\caption{PERs of the proposed algorithm and the algorithm from~\cite{bernier2020low}
			for uncoded LoRa with $\textrm{SF} = 8$ and $N_P = 28$ payload symbols.
            The thick gray line shows the PER for a perfectly synchronized receiver.}
	\label{fig:per-comparison}
\end{figure}

The setup used in Fig.~11 of~\cite{bernier2020low} is however not realistic since, in practice, the value of the fractional STO is uniformly distributed
in the range $\mathopen]-0.5, 0.5\mathclose]$.
We now evaluate the performance by including a fractional STO in the simulations.
Fig.~\ref{fig:per-comparison} shows the overall packet error rate (PER) for our synchronization scheme and the algorithm from~\cite{bernier2020low}.
The transmitted packets again use $\mathrm{SF} = 8$ and are uncoded.
The overall PER (solid lines) is obtained by applying the synchronization algorithm (either the proposed or the one in~\cite{bernier2020low}) to the received packets, estimating and correcting all the time and frequency offsets as each algorithm determines, and finally deciding on the survival of the packet by checking if the symbols of the payload were demodulated correctly.
We observe that the algorithm of~\cite{bernier2020low}, which has not been evaluated before in the presence of a fractional STO,
results in a degraded overall PER performance (solid orange curve) under this more realistic simulation setup.
On the contrary, we observe that the PER performance of the proposed algorithm (solid blue line) is very good and close to the performance of a perfectly synchronized receiver (thick gray line). For example, to attain an overall PER of $10^{-2}$, we observe that our receiver requires at most~$0.5$~dB higher SNR than a perfectly synchronized receiver.

To provide an in-depth understanding of the performance of both algorithms, we deconstruct the overall packet error rate into two error events.
The first error event corresponds to a synchronization failure which results in a complete packet loss. This happens when, after estimation and correction, the residual offsets prevent the receiver from correctly demodulating any symbol in the payload, even in the absence of noise~\cite{ghanaatian2019lora}. The second error event, i.e., a payload error, arises when the receiver manages to correctly synchronize to the packet, but experiences demodulation errors in the payload due to small residual offsets and AWGN.
We note that the overall PER shown in Fig.~\ref{fig:per-comparison} for both algorithms is agnostic to the definition of these two error events.

Regarding the first error event, we consider that the algorithm is effectively unable to synchronize to a packet if,
after the estimation and correction of all four integer and fractional offsets, the demodulated symbols obtained using~\eqref{eq:demod} are all wrong even in the absence of AWGN.
This happens when $|\widehat{L}_{\text{CFO}} + \widehat{\lambda}_{\text{CFO}} - \tfrac{N \Delta f_c}{B}  + \widehat{L}_{\text{STO}} + \widehat{\lambda}_{\text{STO}} - B \tau| > 0.5$,
since a residual total offset greater than $0.5$, i.e., the middle point between two samples, shifts the position of a symbol $s$ in frequency to a DFT bin index other than $s$.
A shift greater than $0.5$ thus prevents the receiver from correctly demodulating any symbol in the payload, even in very high SNR regions or in the total absence of noise~\cite{ghanaatian2019lora}.
We note that the above definition, which treats time and frequency offsets in a unified manner, inherently takes into account the case where a residual integer CFO compensates
for a residual integer STO. Such a case results, in the absence of AWGN, in a correct demodulation of the payload and is hence not considered as a synchronization failure.
It however introduces a small inter-symbol interference (ISI) which degrades the performance during the demodulation of the payload~\cite{ghanaatian2019lora}.

The second error event concerns demodulation errors in the payload that are caused by AWGN.
As explained in Section~\ref{sec:STO}, residual fractional offsets left after the synchronization stage increase the probability of demodulation errors.
Let $\lambda_{\text{res}} = \widehat{L}_{\text{CFO}} + \widehat{\lambda}_{\text{CFO}} - \tfrac{N \Delta f_c}{B} + \widehat{L}_{\text{STO}} + \widehat{\lambda}_{\text{STO}} - B \tau$
be the total residual fractional offset that remains after the synchronization stage, where $\lambda_{\text{res}}<0.5$. When both integer offsets are successfully corrected,
and following~\eqref{eq:cfo-sto-freq}, the DFT of a modulated symbol $s$ becomes
$Y_k = N \cdot \Pi(k - s, \lambda_{\text{res}}) + W_k$, i.e., the impact on the demodulation of both residual fractional offsets becomes indeed equivalent and equal
to $\lambda_{\text{res}}$.
At a given SNR and spreading factor, the payload error probability is strongly influenced by the distribution of the residual values $\lambda_{\text{res}}$ left by the synchronization stage.

In Fig.~\ref{fig:per-comparison}, we observe that the synchronization failure probability is limiting the performance of the algorithm of~\cite{bernier2020low},
especially in high SNR regions. This is because~\cite{bernier2020low} does not correct the fractional STO during the preamble, but only during the payload.
The uncorrected fractional STO increases the probability of wrongly demodulating the preamble symbols, and thus of incorrectly estimating the integer offsets.
On the contrary, our proposed algorithm, which estimates and corrects the fractional STO during the preamble, accurately estimates both integer offsets,
even for low SNR values, resulting in a synchronization failure probability that does not limit the overall performance.

Moreover, the receiver from~\cite{bernier2020low} also exhibits higher payload error probabilities than our receiver.
This stems from the fact that the function used by the estimator of the fractional STO
has a horizontal inflection point at the origin~\cite{seller2018patent}. As such, the inverse function has a vertical tangent at the origin
(see also Fig.~9 of~\cite{bernier2020low}) that increases the variance when estimating very small offsets, leading therefore to instabilities in the estimation of the fractional STO in the tracking loop~\cite{seller2018patent}.
On the other hand, thanks to the low-variance estimator $\widehat{\lambda}_{\text{STO}}$ of~\eqref{eq:estimator-lambda} that is using three upchirps in the preamble,
our receiver is capable of accurately correcting the fractional STO, a fact that results in much improved payload error probability.

\begin{figure}[t]
    \centering
    \setlength\fwidth{0.6\linewidth}
    \setlength\fheight{6.4cm}
    \includegraphics{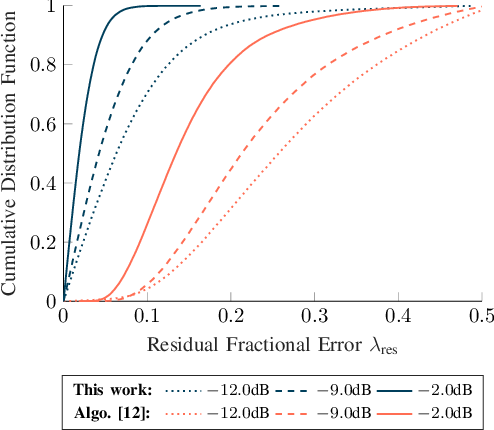}
    \caption{Cumulative Distribution Function of the residual fractional error $\lambda_{\text{res}}$ for the proposed algorithm and the algorithm from~\cite{bernier2020low}, with $\mathrm{SF} = 8$ and $N_P = 28$. Only correctly synchronized frames are included in the metric.
    }
    \label{fig:cdf}
\end{figure}

To better assess the impact of the residual fractional offsets on the payload error probability, in Fig.~\ref{fig:cdf} we plot
the cumulative distribution function (CDF) of the worst-case $\lambda_{\text{res}}$ value for both algorithms. We note that the estimator used in the proposed algorithm results in the same value (which can also be considered worst) for all 28 symbols, since the estimation is performed during the preamble.
For the algorithm of~\cite{bernier2020low}, we do not consider the values $\lambda_{\text{res}}$
after the first demodulation error, to avoid affecting the CDF by the error propagation in the tracking loop (the estimation in the tracking loop may become completely unstable when the DFT bin of a symbol is not correctly identified).
Since both algorithms use the same estimator for the fractional CFO, the differences between the CDFs
mainly illustrate the differences of residual errors on the fractional STO.
For the receiver of~\cite{bernier2020low} and an SNR of $-9$~dB, $60$\% of the packets contain
at least one payload symbol that experiences a residual offset $\lambda_{\text{res}}$ greater than $0.2$.
Such high values of $\lambda_{\text{res}}$ significantly increase the probability of a wrong demodulation of a symbol.
Regarding our proposed receiver, we observe that for the same SNR more than 95\% of the synchronized packets
experience a very low residual fractional error $\lambda_{\text{res}}$ that is below $0.1$ during the demodulation of the payload.
As illustrated in Fig.~\ref{fig:ser-cfo}, such very low fractional offsets have small impact on the demodulation performance.

All the above results strongly illustrate the need for the receiver to correct the fractional offsets before estimating the integer offsets, as well as
the need to use an estimator for the fractional STO with a small variance.
Overall, the performance evaluation indicates that both the early correction of the fractional CFO and STO and the use of the more accurate estimator $\widehat{\lambda}_{\text{STO}}$ instead of $\widehat{\lambda}_{\overline{\text{STO}}}$
enable our proposed receiver to attain PERs of $10^{-3}$ in low SNR regions that are relevant for LoRa.
On the contrary, the receiver from~\cite{bernier2020low} levels off at an error floor for the overall PER that is between $10^{-1}$ and $10^{-2}$.

\subsection{Impact of the CFO Range on the Performance}

As explained in Section \ref{sec:receiver}, the proposed algorithm iteratively estimates the fractional STO in two steps.
The first estimate $\tilde{\lambda}_{\text{STO}}$ is computed with the approximation $\widetilde{M} =  N - \tilde{s}_{\text{up}}$ instead of $\widehat{M} = N - \widehat{L}_{\text{STO}}$.
In practical SNR regions, we have $\tilde{s}_{\text{up}} \approx L_{\text{CFO}} + L_{\text{STO}}$ and the accuracy of the approximation $\widetilde{M}$ thus depends on the value of $L_{\text{CFO}}$.
Large values of $L_{\text{CFO}}$ decrease the precision of $\tilde{\lambda}_{\text{STO}}$, which in turn increases the probability of wrongly estimating the integer offsets $L_{\text{CFO}}$
and $L_{\text{STO}}$ and of experiencing synchronization failures.
Unlike our receiver, the receiver from~\cite{bernier2020low} does not use a similar approximation, its synchronization failure probabilities hence do not depend on
the value of $L_{\text{CFO}}$ when $\Delta f_c \in \left[-\tfrac{B}{4}, \tfrac{B}{4}\right]$.

\begin{figure}[t]
    \centering
    \setlength\fwidth{0.8\linewidth}
    \setlength\fheight{8cm}
    \includegraphics{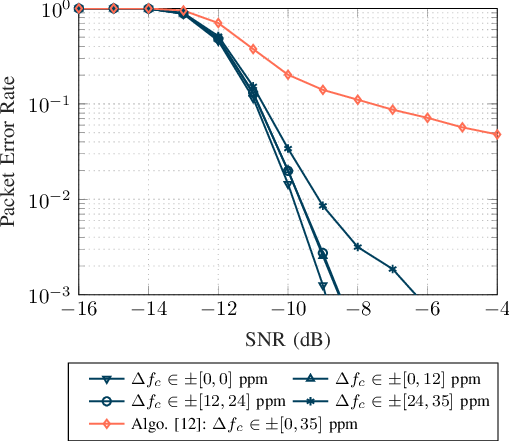}
    \caption{
        PER of the proposed synchronization algorithm for uncoded LoRa with different CFO ranges,
        $B = 125$~kHz, $f_c = 868$~MHz, $\mathrm{SF} = 8$ and $N_P = 28$.
        The PER of the receiver from~\cite{bernier2020low} does not depend on the CFO range.
    }
    \label{fig:per-cfo}
\end{figure}

In Fig.~\ref{fig:per-cfo} we show the overall PER of the proposed synchronization algorithm for $\mathrm{SF} = 8$ and different ranges of CFO.
We observe that CFO values below $24$~ppm only induce a negligible deterioration of the PER.
However, obtaining a PER of $10^{-2}$ with CFO in the range $[24, 35]$~ppm requires $1$~dB stronger SNR than for $\Delta f_c = 0$.
This loss increases to $2.5$~dB for a PER target of $10^{-3}$.
Nevertheless, such large CFO values are unlikely to arise in practice as $20$~ppm is the expected maximum offset for oscillators in commercial receivers~\cite{seller2018patent}.

\subsection{Packet Error Rates for Coded LoRa}

\begin{figure}[t]
    \centering
    \setlength\fwidth{0.8\linewidth}
    \setlength\fheight{7.75cm}
    \includegraphics{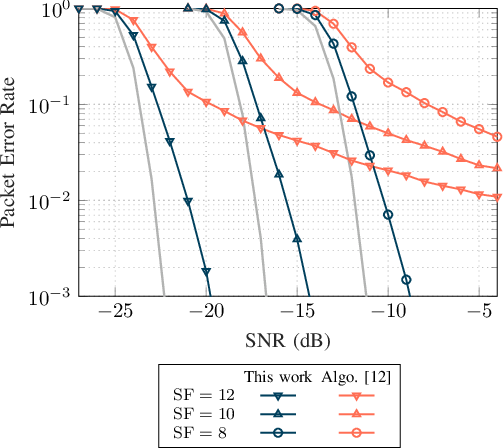}
    \caption{PER of the proposed algorithm and the algorithm from~\cite{bernier2020low} for coded LoRa with different spreading factors, $N_P = 28$ and a Hamming code $(4,7)$.
    The thick gray lines show the PER for a perfectly synchronized receiver.}
    \label{fig:per-sf}
\end{figure}

Fig.~\ref{fig:per-sf} presents the overall PER of the proposed synchronization algorithm and the algorithm from~\cite{bernier2020low} for three different spreading factors
when a Hamming code with coding rate $4/7$ is used.
Regarding first the receiver of~\cite{bernier2020low}, due to the predominance of synchronization failures as described previously,
its error rates level off for all spreading factors at error floors between $10^{-1}$ and $10^{-2}$.
We do not observe any gain due to coding gain for the overall PER of~\cite{bernier2020low}, since the performance in that case is limited by the synchronization
failures in the preamble, which is uncoded.

On the contrary, our proposed receiver does benefit from the Hamming code, especially in low SNR regions.
The decoding stage indeed decreases the probability of experiencing payload errors~\cite{afisiadis2019error}, which are the dominating error events for our receiver at low SNR.
For a target PER of $10^{-1}$ and $\textrm{SF} = 8$, we observe a $1$~dB gain between the coded scenario from Fig.~\ref{fig:per-sf} and
the uncoded scenario showed in Fig.~\ref{fig:per-comparison}. This gain however decreases at higher SNRs as synchronization failures
tend to become the dominating error events.
Whereas our receiver requires only $1$~dB higher SNR than a theoretical perfectly synchronized receiver to reach a $10^{-3}$ PER in the absence of coding,
this difference increases to $2$~dB with coding. In the latter scenario, only synchronization failures significantly cause packet errors,
whereas in the uncoded case, the probabilities of experiencing a synchronization failure or a payload error are almost identical, as shown in Fig.~\ref{fig:per-comparison}.

\section{Complexity analysis}
\label{sec:complexity}

We finally discuss the complexity of our proposed synchronization algorithm and the algorithm of~\cite{bernier2020low}.
Both algorithms rely on the conventional demodulation stage of a LoRa receiver, with some additional computations and memory requirements.
All synchronization computations are performed at the Nyquist-rate $f_S = B$, as explained in Section~\ref{sec:simu}.
We show that the overhead of the synchronization is minor compared to the complexity of the demodulation stage itself.

The proposed algorithm and the algorithm from~\cite{bernier2020low} use similar estimators for the fractional CFO and for the integer offsets.
Estimating the integer offsets $L_{\text{CFO}}$ and $L_{\text{STO}}$ involves only three simple arithmetic operations.
The estimator $\widehat{\lambda}_{\text{CFO}}$ requires a single $\angleF$ computation and a constant, but small, number of complex Multiply and Accumulate (MAC) operations, as listed in Table~\ref{tab:complexity}.
Regarding the fractional STO, both our estimator $\widehat{\lambda}_{\text{STO}}$ from~\eqref{eq:estimator-lambda}
and the estimator $\widehat{\lambda}_{\overline{\text{STO}}}$ from~\eqref{eq:lambda-hat-bernier} carry out a fixed number
of arithmetic operations on three DFT outputs $Y_i$, including one complex division. It is worth underlining that the latter estimator requires an inversion of the non-linear function $T(x)$, whereas the former does not.

\begin{table}[ht!]
    \centering
    \begin{tabular}{|c|c|c|}
        \hline
        \textbf{Offset} & \textbf{Operations} \\
        \hline
        $L_{\text{CFO}}$, $L_{\text{STO}}$ & Two integer additions, one integer division \\
        \hline
        $\lambda_{\text{CFO}}$ & 11 complex MACs, 1 $\angleF$ \\
        \hline
        $\lambda_{\text{STO}}$ & \begin{tabular}{@{}c@{}} 6 complex additions, 10 complex MACs\\ 2 complex divisions\end{tabular}  \\
        \hline
    \end{tabular}
    \caption{Operations carried out by the proposed algorithm per packet to estimate the integer and fractional parts of the STO and CFO.}
    \label{tab:complexity}
\end{table}

The algorithmic complexity of the demodulation stage given in~\eqref{eq:demod} is $\mathcal{O}(N \log{} N)$
when the DFT is implemented using a Fast Fourier Transform (FFT), and hence depends on the spreading factor.
This stage also requires the storage of $N$ samples in memory, since the dechirping operation and FFT can be performed in place.
As the studied synchronization algorithms involve only a constant and limited number of operations, their computational cost is negligible compared to the complexity of the demodulation stage,
especially for large spreading factors.
Our proposed algorithm exhibits a very low computational overhead, similar to~\cite{bernier2020low}, but much better performance results, making it a suitable candidate for low-power IoT end nodes.

\section{Conclusion}
In this work, we design and evaluate a LoRa synchronization algorithm robust to carrier frequency and sampling time offsets.
We derive a new accurate estimator for the fractional STO, as a precise correction of this offset is crucial to reach the low sensitivity levels
provided by the LoRa modulation.
We also show that the estimation of this fractional component is intertwined with the estimation of the integer parts of the STO and CFO.
To avoid a complex joint estimation of these offsets, we propose instead a low-complexity synchronization algorithm that iteratively estimates the fractional STO
using the new estimator.
For a target packet error rate of $10^{-3}$, performance evaluations show that a receiver using the proposed synchronization algorithm requires only $1$~or $2$~dB higher SNR compared to
a perfectly synchronized receiver, while requiring only negligible computational complexity.

\section*{Acknowledgments}

The authors would like to thank Prof. A. Burg, Prof. A. Balatsoukas-Stimming, and Dr. R. Ghanaatian for useful
discussions on the LoRa PHY.

\bibliographystyle{IEEEtran}
\bibliography{IEEEabrv,paper}

\end{document}